**Almost-anywhere Theories**

**Reductionism and Universality of Emergence**

Ignazio Licata

Institute for Scientific Methodology, Palermo, Italy

Ignazio.licata@ejtp.info

We aim here to show that reductionism and emergence play a complementary role in understanding natural processes and in the dynamics of science explanation. In particular, we will show that the renormalization group – one of the most refined tool of Theoretical Physics – allows understanding the importance of emergent processes' role in Nature indentifying them as universal organization processes, which is to say they are scale-independent. We can use the syntaxes of Quantum Field Theory (QFT) and processes of Spontaneous Symmetry Breaking as a trans-disciplinary theoretical scenario for many other forms of complexity, especially the biological and cognitive ones.

## 1. The false opposition in the images of science

The development of scientific thought is commonly and by now quite unsatisfactory regarded as oscillating between a linear and thus "normal" course of science and quick catastrophes leading to a radical paradigm shifting (Godfrey-Smith, 2003). Maybe, the idea of a paradigm shifting could make more sense in the century when Relativity, Quantum Physics and Molecular Biology were born in the span of few decades. Watching more carefully contemporary research reveals that normal science is instead a "coarse-grain" and long term approximation. We can actually see a relentless fluctuating of research programs with a temporary prevailing of one among the others, so becoming the order parameter on the knowledge developing axis. It is worthy noticing that the best description of the evolution of scientific knowledge can always be drawn from the internal concepts of science: the old images mirror the long development of linear Classical Physics, whereas the idea of always dynamically competing research programs derives from the most recent Statistical Physics of non-linear processes and Quantum dissipation.

On a more rooted level, there are the *metaphors* of science which provides the research programs with the proper philosophical *humus*, so to define their own epistemological directrixes. The *Cosmic Code* (Pagels, 1984; see also Penrose, 2007) is the first and more ancient example. Essentially, the basic assumption is that the more the scientific research widens its methodological tools the more it will be able to grasp the set of "fundamental laws" existing "out there". Once such basic laws are individuated, it will make possible to describe any Nature's manifestation as a necessary consequence of a small set of propositions representing the universe *a priori*. That would crown the ambition for the "Theory of Everything", which is – in its different mathematical forms – the modern equivalent of that tending towards the first principle typical of Greek philosophy; further evidence that human thought – rather similarly to biological organisms - is a continuous variation on few themes.

Such image of science is quite "architectural": a tower of the world's descriptions grounding on a single fundamental Theory of Everything and rising to upper levels only by the strength of logical inference. That of a *chain of theories* is the most proper way to represent the role of reductionism in science. Recently, a new image has been making its way within the *sciences of complexity*: the *Emergence Theory* that has often been regarded as the exact opposite of naïve reductionism, which is naively described as well. Reductionism is centered on the analytical identification of "the bricks making the world up" or the constituting units of the system to explain its features, whereas emergence stresses on irreducibility of collective behaviours to their elementary constituents. Such vulgate makes reductionism and emergence look alike, because them both states two different kind of obvious.

We aim here to show that reductionism and emergence play a complementary role in understanding natural processes and in the dynamics of science explanation. In particular, it is just when we use emergence that we can actually realize *why* the world description appears as a hierarchy of levels of organization on different scales, thus deeply grounding the theoretical tower where the schemes are piled one on the other so that each scheme is the basis for the next one. In order to understand such dynamical complementarity of the two reductionism/emergence key-images it is necessary to demolish both the naïve versions and providing a more conceptually defined one. We will show that the most natural theoretical context to set such program is the general logic frame of the Theory of Quantum States. In particular, we will show that the renormalization group – one of the most refined tool of Theoretical Physics – allows understanding the importance of emergent processes' role in Nature indentifying them as universal organization processes, which is to say they are scale-independent. Far from being something only related to many-constituent macroscopic systems, emergence plays a decisive role even in those theories we usually regard as "fundamental", thus opening new perspectives on frontier problems of theoretical physics. Viceversa, we can use the syntaxes of Quantum Field Theory (QFT) and processes of Spontaneous Symmetry Breaking as a trans-disciplinary theoretical scenario for many other forms of complexity, especially the biological and cognitive ones.

2. **Complexity: a simple introduction**

Let us consider a system S and a set of measurements M relative to S, and put as $T_i$ the set of theories describing the different system's features. Each theory of the set is related to a set of observational data O gained by an apparatus M. It is worthy noticing that the relation between the theoretical *corpus* and the measurement apparatuses is never banal: a change in M can lead to new observables and thus to new theoretical exigencies or to confirm an already existing theory in a finer way, whereas different theories can make reference to the same observational set.

It is typical of reductionism stating that the following program is suitable for any physical system S:

a) $T_i$ can be organized according to a logical sequence of the kind $T_{i-1} \prec T_i$, where the symbol $\prec$ stays for "physically weaker than". This means that the $i^{th}$ theory brings more information than the previous one, and so – by using the fit mathematical tools – the preceding theories of the sequence can be derived from the strongest one;

b) There exists a final $T_f$ which exhaustively, completely and coherently describes any features of S. So, describing the "everything theory" of S means that it is possible to perform the maximal algorithm compression of the information (Chaitin, 2007) extracted from S by measurements M and each observables O find their place within the theoretical chain $T_{i-1} \prec T_i$ as well as within the final theory $T_f$. The knowledge of S provided by the final theory can be considered as the fundamental explicative level, and the $T_i$s as phenomenological descriptions, a sort of limit cases approximating the final theory.

Such idea brings inside a misleading simplicity tending to self-referentiality, which is also the basis for its reputation. FAPP (*For All Practical Pourpose*) [1], this is the messages shortly conveyed by "Particle Physics": according to a well-known Steven Weinberg saying, the most useful scientific descriptions "point the conceptual arrows down", towards the elementary constituents of matter. At fundamental level, there are just a fistful of fermions and bosons, the connections between the interactions intertwining them and, at least, some global constraints of cosmological nature. All the rest can be included in the renowned category "the phenomenon Y is nothing but X", even if the logic reduction process can be so complicated to be classified as a computational catastrophe.

---

[1] Such expression is borrowed from John Bell (1928 – 1990), one of the sharpest investigators of Quantum Mechanics conceptual foundations. By the acronym FAPP – For All Practical Purpose – he always reminded himself and the community of physicists that the usefulness of an alternative interpretation of the theory had to come to terms with the fact that the standard QM interpretation worked well in the most varied applications.

A more careful examination is enough to become convinced that such kind of reductionism is totally unjustified and require accepting a lot of obscure postulates. In short, we have concepts finding their plausibility in Newtonian mechanics which anachronistically migrate into a radically and irreversibly non-classical vision. In particular, a crucial problem, not solvable by automatic procedure, is nonchalantly skipped: providing a complete and exhaustive system description by a theoretical chain of the kind $T_{i-1} \prec T_i$ imply choosing the significant information characterizing the $i^{th}$ theory, and consequently the information into play is supposed to be homogeneous, syntactically univocal, independent from the observer's choices and unaffected by measurements throughout the conceptual path of the reductionist program.

To put it in straight physical words, this is the equivalent of stating that the knowledge of an energetic range univocally fixes the organizational forms which can be detected at that level. Being able "to saturate" the description of a system S with a single theoretical apparatus $T_f$ means that information about any system' states can be extracted by means of a finite – or, at least, countably infinite - sequence of M measurements. By following an operational criterion, these ones can be easily codified by a Turing Machine, a robot we shall name Turing-Observer which has been tailored on $T_f$ rules. It can be proved that such kind of program can be carried out – on a very limited range – only when a system is informationally closed with respect to an observer, a system of which we can always track the values of state variables and describe the evolutionary laws through recursive functions (Licata 2006a; 2008a; 2008b). That is the case of some "toy-models" inspired to classical Physics, such as Cellular Automata (CA). As is well-known, Wolfram-Langton classification (Langton, 1991) identifies four fundamental classes of CA which can be ordered according to a sequence at the varying of $\lambda$ - parameter, which corresponds to a sort of generalized energy:

Class I (evolves to a homogeneous state) $\rightarrow$ Class II (evolves to simple periodic or quasi-periodic
patterns) $\rightarrow$ Class IV (yields complex patterns of localized structures with very long transients, like in the famous Conway *Life Game*) $\rightarrow$ Class III (yields chaotic a-periodic patterns).

Such classes are the discrete counterpart of well-known systems in the Theory of Continuous Dynamical Systems; it makes the CA extremely powerful simulation tools. The I, II, III classes respectively correspond to information compressing systems (tending to a fixed point, for ex. maximum entropy state), information amplifying systems in polynomial time, such as the famous non-linear and dissipative systems studied by Prigogine and in Synergetics (Prigogine, 1997; Haken, 2004), and structurally unstable chaotic systems. The class IV is a dynamic typology amid unstable systems and dissipative ones.

In a world like that, Everything Theory is represented by the "fundamental equations" made by the number of rules compatible with its topology (dimension, number of states, neighbourhood rules "switching on" and "switching off" a state). The simplest case is that of a two-state one-dimensional CA with neighbourhood rules regarding the cells standing on the right and on the left of the one under consideration, so there are $2^3 = 8$ possible patterns for a neighbourhood of 3 cells and $2^8 = 256$ rules, each in its class. Any pattern emerging in this toy-world can be measured by a Turing-observer and lead back, in principle, to the fundamental rules by a purely computational process. The situation is similar to chess; extremely complex configurations can come out and many pieces are not on the chessboard anymore in middle-game, but no matter how complex the number of configurations can be it is finite and independent from the observer choices and this is what makes possible "retracing" the game starting from the initial position (boundary conditions) and the laws (how-to-move-pieces rules).

Actually, in chess-game as well as in CA, a global situation can be so complex not to be easily connected to the local rules, thus requiring making use of a high level language plus a statistical study of the configurations. Anyway, the lack of a close correlation between local and global predictability – such as in halting problem and chaotic dynamical systems – does not imply the failing of casual determinism connected with the observer possibility of step-by-step following the system. In such toy-world no authentic "newness" occurs, emergence is just the manifestation of patterns obtained by purely computational processes. In this sense, information is homogeneous, syntactically defined and unaffected by the observer choices and measurements as well.

In conclusion, *a genuinely reductionist program can only take place in a system that is informationally closed with respect to the observer, and its evolution can be described as a Shannon-Turing intrinsic computation.*

The central role of observational choices and the not strictly algorithmic nature of relationships between the different descriptive levels in radical/observational emergence have been investigated in Hyperstructure Theory by Baas and Emmeche (Baas and Emmeche, 1997; see also Cariani, 1991). Let us consider a set of systems $S_1$, the interactions between systems $INT_1$ and a set of observational procedures $M_1$. Observations are peculiar forms of interaction with the set of systems; they detect information and provide the global system properties. The $INT_1$ generate a new structure of the kind $S_2 = \langle S_1, M_1, INT_1 \rangle$, called *emergent structure*, to which a new set of observational procedures $M_2$ can be applied. A property P is said to be emergent if and only if $P \in \langle S_2, M_2 \rangle$. The hierarchy $S_n$ is called a Hyperstructure.

The Baas-Emmeche definition of emergence formally fulfils Philip Anderson's criticism of reductionism expressed by his famous statement "*more is different*" (Anderson, 1979). In fact, the behaviours we observe at emergent level are clearly compatible with those at basic level and yet not merely reducible to them to such a degree they require new observational and theoretical tools.

In sum, as for the logical relationships between descriptive levels, two kind of emergence can be admitted. *Phenomenological/computational emergence* occurs when it is always possible to find a formal and algorithmic relation between the two levels, and *radical/observational emergence* where, on the contrary, no univocal deductive relation able to give evidence of such connection can be drawn. Thus, radical emergence "breaks" the theoretical chain of naïve reductionism and makes possible to study the different organizational levels of a system (as separate items) in autonomous way, independently from the analysis of its constituents. The possibility to do this is one of the most extraordinary phenomenons we can run into in science and has actually favoured the historical development of research fields connected with the *investigation of scales*. On each scale we have found "elementary units", but what is highly significant is that we have been able to study their organizational forms putting aside almost anything concerning the inferior or superior scales. A noticeable example is the developing of classical physics, which nowadays has been proved to be a refined "compromise" of quantum effects.[2] On the one hand, we know that an essential truth lies in reductionism – nervous system can be studied as a complex set of neuronal networks, biological systems are made of cells, molecules consists of atoms, which in turn consists of nuclear matter and so on up to the theory of elementary particles – on the other, emergence shows us the possibility to study Nature on extremely different and far organizational scales without worrying too much of the microscopic details related to the lower level.

Thus, grasping the complexity of the world requires a theoretical scenario which reconciles the *quantum veritatis* of reductionism with radical emergence processes. The key starting point to understand such scenario is that the world is not made of cellular automata, chess pieces or

---

[2] There are different approaches to the emerging of classical world from quantum one. As for de-coherence, see Griffiths, 1996; there are at least two theories deriving from Bohm: see – on the one hand – the "classical limit" by Allori et al., and – on the other hand - the switching from non-local to local information in Hiley, 2000. A particularly interesting and effective approach, based upon Quantum Field Theory, is that of Vitiello, 2005, which describes the emerging of classical Physics as a dissipative quantum effect.

Newtonian particles, but it is fundamentally quantum-based, informationally open to the observer (entanglement and non-local information) and affected by measurements.

3. **Quantum variations on a Mexican hat**

Let us put here aside the mathematical details and take into consideration a potential called – patently for its shape – *Mexican hat*. This kind of potential can be found on many different scales, from particle physics to condensed matter. It is related to a lot of "interesting" situations from organizational viewpoint.

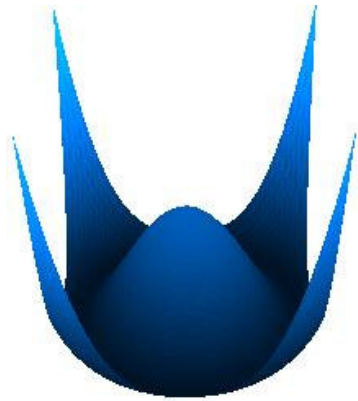

Fig. 1- "Mexican hat" potential

Let us consider a marble standing in unstable equilibrium on the top of the sombrero. When a slight variation occurs so breaking the equilibrium, the marble will roll along the slant down to some position in the circular valley. The global structure of the dynamic situation obeys to the general symmetry principles (the Mexican hat does not change its shape), but the final state is highly asymmetrical. As a classic case, it is a greatly banal situation, but if the marble under consideration is an infinite state quantum system and the sombrero is the potential defining its dynamic-evolutionary possibilities, the whole matter becomes quite interesting. The rolling down is really pertinent an image for radical emergence phenomena in Quantum Field Theory (QFT) (for an essential and brilliant exposition of the theory see A. Zee, 2003).

When one of the parameter linked to the available energy changes, the system will distribute in one of the many possible *ground states*, with a consequent energy redistribution characterizing its macroscopic properties. Each "marble position" expresses a different energy arrangement of the system and, differently from the classical case, there is no possibility to forecast any detail about the final state; because the renowned quantum dice which Einstein was so worried of are not informationally closed with respect to the observer, whereas the statistics of quantum objects – Fermi-Dirac for fermions and Bose-Einstein for bosons – are radically different from classical statistics and provide a rich phenomenology of organized states.

Just to be more precise. The key idea is that in infinite state quantum systems different and not unitarily equivalent representations of the same system are possible, and consequently phase transitions structurally modifying the system, too. This occurs by means of the *Spontaneous Symmetry Breaking* (SSB), i.e. a process which does not let all the states compatible with a given energy value invariant. What usually happens is that when a given parameter varies, the system will settle on one of the possible fundamental states, so breaking the symmetry. This brings to a balancing how it is shown by the emergence of long-range correlations associated with Higgs-Goldstone bosons, which act to make the new configuration stable. The boson condensed states can be fully considered as forms of *macroscopic coherence* of the system, and they are peculiar to the quantum statistics formally depending on *indistinguishability of states* with respect to observer. The new system's phase requires a new description level for its behaviours, so we can speak of radical emergence. Many behaviours of great interest in Physics on different scales are included within SSB processes, such as phonons in a crystal, Cooper pairs in superconductivity phenomena, Higgs mechanism and multiple vacuum states in elementary particle physics, inflation and formation of the "cosmic landscape" in Quantum Cosmology. It is so reasonable to suppose that the fundamental processes for the formation of structures essentially and critically depend on SSB and the QFT "syntax" makes possible to grasp them.

A question of great interest comes out when comparing the "ideal model" of emergence proposed by the language of dense quantum systems with the more classical, traditional and "not-classical" ones of Prigogine dissipative systems the self-organization processes on the edge of order and disorder (Pessa, 2002). Such problem is strongly correlated to the emergence of classical world from quantum one, and a promising approach is to consider the traditional – classical or semi-classical and critically depending on opportune boundary conditions – self-organization theories as emergent residual "traces" of SSB processes. The most of complex systems we deal with have after all a finite dimension and a very high, but not infinite number of degrees of freedom. An answer could be that these systems are the outcome of a "freezing" of the degrees of freedom typical of the SSB system and *all* the classical self-organization phenomena are the consequence of quantum process of symmetry breaking (see Wadati et al., 1978a, 1978b; Umezawa, 1993; Anderson and Stein, 1985; Kuma and Tasaki, 1994; Pessa, 2008). In general, phenomenological emergence manifestations are a particular case of quantum radical emergence.

How can SSB radical emergence be compared to the phenomenological detection of patterns and which are instead the radically quantum features? In SSB processes, the phase transition is likewise led towards a globally predictable state by an order parameter, i.e. we know that there exists a critical value beyond which the system will find a new state and exhibit macroscopic correlations, and here too a relevant role is played by boundary conditions (all in all, a phonon is the dynamic emergence occurring within a crystal lattice and it does not make any sense out of it). Moreover, in SSB there exists an "adjustment" transient phase whose description is widely classical. Where the analogy fails and we can actually speak of an irreducibly non-classical feature is the bosonic condensation, which is a non-local phenomenon. In a classical dissipative system we can, in principle, obtain information on the "fine details" of bifurcation and know where the marble will fall, whereas in SSB process it is not possible because of the very nature of the *quantum roulette*!

The problem of statistical mentalics - just to say it with the famous expression of Douglas Hofstadter (Hofstadter, 1996 ) – is a quite interesting one, which is to say the idea to deduce the symbolic structures of cognitive processes from the sub-symbolic organizational processes on neural level. Smolensky (Smolensky, 2006) showed that within the traditional formalism of connectionism this program can be realized only in few extremely simplified cases. A totally different approach is that of Dissipative Quantum Brain, where the dynamics of the collective state variables are shaped on dense quantum processes and the possibility to exhibit SSB phenomena as well. That is the case when the "statistical mentalics" program can be realized, and the Quantum Field Theory substantially plays the role of a "super-neural net" able to exhibit radical emergence processes triggered by the system/environment relations which strikingly correspond to the observed in laboratory functional structures (for a general introduction see Vitiello, 2001 and Licata, 2008a; the classical and exemplary clear paper is Umezawa-Ricciardi, 1967; two more technical references are Vitiello and Freeman, 2008; Vitiello and Pessa, 2004).

4. **Universality, Emergence and the Renormalization Group**

Strictly speaking, in QFT similarly and even more radically than in QM a particle is not a nomological fundamental "object", but an event fixed by a network of relations whose conditions of existence are set by the dynamics of the interacting fundamental fields – called Heisenberg fields -, the correlations between the energy levels given by quantum statistics, and the emergent dynamics of the phenomenological fields. The emergent dynamics are directly related to the observed objects, such as particles, which are formally described as asymptotic states of field. What an "elementary object" is depends on the scale under consideration (energy, length, times). How many phenomenological levels do exist? How far –upward and downward - can the "zoo" of fundamental objects related to each level reach? Reductionism comes up again from these questions, but fortunately the QFT formal structure makes possible to arrange the matter of the relations between different descriptive levels clearly, so avoiding the *recursion ad infinitum* suggested by the typically reductionist image of the matryoshka-like particles.

As often it has happened in the history of science, the original solution to this problem comes out from the necessity "to embank" the infinites occurring in the theory by an *ad hoc* procedure called renormalization.

The status of renormalization as an unsatisfying heuristic tool spans from '30 to '80, when Kenneth Wilson rigorous formulation of the Renormalization group (for an introduction see Wilson, 1997) provided the theory with a new physical meaning, so decisively contributing in delineating a new virtuous relation between reductionism and emergence. The theory is a powerful self-consistence condition on "effective" field theories (EFT, Effective Field Theory), a mathematical mechanism to individuate the "correct" and physically "stable" phenomenological scales. The renormalization group functions just as a mathematical zoom lens which allows looking a physical system with different resolution degrees, mediating on the peculiar properties characterizing each scale and gaining the significant features and invariances which occur during the passage from a phase to another in a quantum system.

If we define the phenomenology related to a given range of energies and masses as a description level, it will be possible to make use of the renormalization group (RG) as a *resolution tool* to pass from a level to another by varying the group's parameters. So, we obtain a succession of descriptive levels, a tower of *Effective Field Theories* (EFTs), each with a given *cut-off*, able to grasp the peculiar features of the investigated level. In this way, each level is linked to the others by a rescaling of the kind $\Lambda_0 \to \Lambda(\sigma) = \sigma \Lambda_0$, where $\Lambda_0$ is the *cut-off* parameter relative to a fixed scale of energies/masses into play that defines a single level. Each level is characterized by a coupling parameter which defines the interaction and organization among the objects of a specific phase. By using an analogy not so far from the actual mathematical features of the theory, the situations is similar to that described by the fractal theory when trying to grasp, within limits, the recurrence of analogous structures at different scales (Lesne, 1998).

The universality of SSB mechanism is deeply linked to such aspect, and using the QFT formalism as the general theory of emergence is based exactly on such powerful condition of theoretical coherence. In fact, the possibility itself to connect different levels through the renormalization procedure is the unequivocal sign that *Nature plays the quantum emergence game on different scales until the emergence of the classical world.* Robert Laughlin (Laughlin, 2006; Laughlin and Pines, 1999) called "laws of protection" such Quantum Theory features.

After an energy scale has been fixed, we can study it without worrying too much of the properties of the inferior level constituents just thanks to the constructive richness of quantum statistics and the universal structure of SSB processes focalized by the renormalization group. *In a more radical way we can say that it is just the QFT which allows, on each level, to generate interacting objects according to similar organization schemes, by individuating the "constituents" of the system on each scale.* In this sense the QFT is at the same time a TOE (Theory of Everything) and a TOO (Theory of Organization). On the other hand, QFT does not lack many internal limitations. One is the necessity to postulate some fundamental characteristics, such as the properties of fermions and bosons; we could thus ask if the tower of EFTs is infinite and especially if there may be found an even more fundamental theory lying at the bottom that has not the form of QFT regulating the behaviour of far more exotic objects like it happens in the brane theory or loop theory.

So we find the matter of reductionism within a new context, but it is just the extension of the "laws of protection" which suggests some way to approach the problem so subtly eluding naïve temptations.

## 5. Ideas: old like brand-new ones, new and revolutionary

Essentially, the history of the creation of structures in the Universe is a succession of phase transitions led by SSB processes until the self-organizing morphogenesis appearing of the classical world. At this point the temptation to introduce some fundamental "substance" from which the known forms of matter and energy emerge strongly arises. The ideal candidate is the quantum vacuum - the modern heir of the old aether – whose stochastic fluctuations are what more or less is needed to make it the effective "ultimate" support for the structure of QFT. The idea has come out again and again in the last years and has tempted top scholars such as Werner Heisenberg, Andrei Sacharov and David Bohm (see Genz, 2001). One of the key problem is the necessity to extract the statistics of bosons and fermions without introducing any parameter but those derived by the fundamental principles of the theory; this is actually the main limit of QFT and gauge theories of interactions derived by its formal apparatus, such as the standard model, where the values of such measurements are derived by experiments and "manually" put in the mathematical model.

A more recent line of research is that of Super-strings or Branes originating from the extension of some '60/'70 models of strong interaction. It is there postulated a mathematically very elegant dynamics of multi-dimensional objects from which space, time and the "ordinary" (but it is a matter of scale!) forms of matter emerge, except for some problems with the excessive mathematical "powerfulness" of the theory, the "hunger for additional dimensions" (Laughlin, 2005) and the difficulty of experimental tests. In order to study the science communication models, the famous Brian Green's "The Elegant Universe" is emblematic (Greene, 2003). The step-by-step magnificent fortune of the theory is praised and the reader's fantasy is delighted for about more than four hundred pages, as for the conceptual flaws of theory they are hastily stuffed in the last chapter! Fortunately, Peter Woit and Lee Smolin counterbalanced that untenable propaganda (Woit, 2007; Smolin, 2007). The Loop Theory derives instead from a formalism Penrose ideated for Quantum Gravity: the spin networks. In this theory, the continuous tiling of space and time emerges from *spin foam* according to a Leibnizian relational logic which inspires the work of Carlo Rovelli ( Rovelli, 2004).

These problems are closely connected to cosmology and quantum information. The former one acts as general boundary condition; and it has been shown that adopting certain global topologies imposes very strong constraints to the possible modifications of QFT. The

DeSitter Universe, recently under heated debate both as fundamental cosmological model and primeval phase of the traditional Big-Bang scenario, cuts off some divergences – technically called infrared problem – and strengthens the consistency of the theory. Another way to study the question is to try to understand how classical local information, that is to say the space-time causal structure, the time arrow and the relations between physics and computation, can emerge from archaic Universe's non-local quantum information (Licata, 2006b, 2008a, 2008b, Chiatti and Licata, 2008). The basic idea is quite simple: classical concepts such as time are not autonomous citizens in any genuinely quantum view, they are instead "enclosed" in the wave-function. The DeSitter Universe global structure as the space of quantum observables makes use of cyclic imaginary time. In order to pass to classical and local notions such as real time, it is used Wick rotation, a traditional mechanism of physical mathematics. In this context, Wick rotation takes up a new physical meaning, because it selects classes of observers where quantum information becomes "condensed" in classical histories on the event space. In this way, it is possible to provide a classical and computational reconstruction of the Universe as forms of emergence from the "archaic" quantum magma.

By developing the Wheeler's "It from the bit" program under quantum context (It from qbit), many authors have shown that scale-free graphs can be detected on different levels. These scale-invariant networks now appears as a fundamental ingredient for any future organization theory and in 2000 Bianconi and Barabasi have demonstrated that Bose-Einstein condensates and phase transitions are formally equivalent to the dynamics of scale-invariant networks (Bianconi and Barabasi, 2000). Also in this case the explanation is conceptually simple: what counts in a phase transition is not the "matter" into play, but organizational processes, and scale-free networks are an effective compromise for information transmission between too rigid connections and extremely fluid ones. Such outcome builds a bridge between the physics of mesoscopic systems and the study of biological, cognitive and social systems (Requardt, 2003; Zizzi, 2008; Licata 2008c; Lella and Licata, 2007, 2008).

After all, there is no need for any fundamental "matter". In the S-Matrix program developed by Geoffrey Chew during the '60/'70 to describe strong interactions, "fundamental objects" had already disappeared to be replaced by a mathematical structure which satisfies some symmetry general conditions. Later, Chew proposed a methodological philosophy, the well-known *bootstrap model,* based on the idea of "nuclear democracy": no particle can be considered as "fundamental", but each one – just to put it in reductionist words – is made of the other ones. The S-matrix history is greatly interesting. For some years it was the main antagonist of QFT and then quickly fell into disfavour, a victim of the complexity of its mathematical description compared to the – apparent! – simplicity of the Theory of Quarks. Lately, the S-matrix ideas have merged into the more ambitious and advanced program of Superstrings and M-Theory. In particular, the idea that the "fundamental theory" core is not any form of pre-matter, but a set of mathematical conditions relative to super-symmetry and self-consistency (Cushing, 2005).

In recent times, Holgar Nielsen – following the principles defined by Laughlin and Pines as connected to SSB characteristics – has proposed a radical vision according to which *the physical laws themselves are emergent phenomena*. According to such hypothesis the Universe we observe is a phase transition, and consequently its structures are largely independent of any fundamental "matter" or "law", in the same way as all the boiling processes are alike independently of the kind of liquid. Taking into account that in this case - as well as any unification program - the "liquid" is at Planck scale; Nielsen has suggested *shifting the research axis from laws to the dynamics of processes*, which is to say focusing on a minimum set of mathematical structures able to describe the generalized universal features of a phase transition (Nielsen et al. 2007). In spite of its being founded on extremely simple concepts, Nielsen's Random Dynamics has been greatly successful in explaining Quantum Mechanics as well as General and Special Relativity, has derived Yang-Mill equations which all the standard model-related unification programs are based upon, and has even found out a general explicative model for many properties of space-time – such as the number of dimensions – and for many features of "elementary" particles. Nielsen program can be outlined as a "conceptual ladder" starting from a mathematical "machinery" which is essential in "switching on the world" up to bringing forth many laws which are still considered as fundamental. In practice, the fact we are in a world ruled by some laws rather than different ones depends on a mechanism similar to that we have observed with the "Mexican hat", and corresponds to the casual choosing of a "world logics" where self-organizational forms occur. Apart from its achievements, Random Dynamics has the merit of explicitly stating an "uncomfortable truth" shared by many research unification programs: Gauge field equations are extremely complex and structurally unstable, therefore the use of non-perturbative exact tools is prohibitive; on the contrary the possible perturbative approaches are almost infinite. It is thus difficult amid this theoretical ocean to find a reliable and univocal theory. As Laughlin wittily points out, seemingly competing unified theories are probably not falsifiable in our current state of knowledge. So, what is going on is a "war of patents" for mathematical technologies and the connected "fundamental" world conceptions rather than a genuine scientific opposition.

Bohm and his collaborators had already studied the algebra of processes by holomovement theory, non-commutative algebras are used to investigate the relationship between quantum implicate order and classical physics' explicate order (Hiley, 2001; Bohm and Hiley, 1995; Monk and Hiley, 1998). The conceptual shifting from a Parmenidean logics of laws to an Eraclitean logics of processes represents the most radical epistemological perspective in Theoretical Physics.

## 6. A lesson of wisdom from "the Middle Way"

Since the age of Galilei, when it first became aware of its cultural autonomy, scientific thinking inherited a push towards the quest for ultimate truth – a fundamental matter ruled by a bunch of essential laws - from philosophical tradition. In its different historical roles and in the deep mutations of its crucial stages, Theoretical Physics – considered for a long time the never rivalled model of science for all the other disciplines -  has always maintained the idea of an ultimate vision centered on fundamental components.

Physics of Emergence represents a deeply different way to look at the world; it is rich of strong implications for understanding of the physical world. As we have seen in our short survey, it is not a conception opposed to reductionism, but its natural counterpart. In addition, when placed within the context of Quantum Theory, emergence provides the effective reasons for the reductionism proper working. *It is the universal nature of emergence on every measurement scale which makes possible to identify on each level some "fundamental*

*constituents and, above all, to study their organizational processes.* The essential core of such lesson does not come from far domains, but from the Physics of "The Middle Way" (Laughlin et al., 2000), which is to say the physics of ferromagnets, superconductors, superfluids, proteins and neural networks; all those areas of Theoretical Physics which have always kept in vital and virtuous contact with the experimental dimension and are still able to surprise us, as it is witnessed by high temperature superconductivity. Even if each level get its own peculiarity, emergence processes acts on any scale, almost anywhere, displaying an extraordinary variety of phenomena, and guiding us towards a knowledge no more centered on the "ultimate" laws, but on the organizational factors of the physical Universe complexity.